\documentclass[aps,prd,superscriptaddress,12pt,showpacs,notitlepage]{revtex4}
\usepackage{amsmath,amsfonts,amssymb}
\usepackage{color}

\newcommand{\be}{\begin{equation}}
\newcommand{\ee}{\end{equation}}
\newcommand{\bea}{\begin{eqnarray}}
\newcommand{\eea}{\end{eqnarray}}

\begin{document}
\title{Symmetries and exact solutions of the BPS Skyrme model}
\author{C. Adam}
\affiliation{Departamento de F\'isica de Part\'iculas, Universidad de Santiago de Compostela and Instituto Galego de F\'isica de Altas Enerxias (IGFAE) E-15782 Santiago de Compostela, Spain}
\author{C.D. Fosco}
\affiliation{Centro At\'omico Bariloche,
Comisi\'on Nacional de Energ\'\i a At\'omica, and  Instituto Balseiro,
Universidad Nacional de Cuyo,
R8402AGP Bariloche, Argentina}
\author{J.M. Queiruga}
\affiliation{Departamento de F\'isica de Part\'iculas, Universidad de Santiago de Compostela and Instituto Galego de F\'isica de Altas Enerxias (IGFAE) E-15782 Santiago de Compostela, Spain}
\author{J. Sanchez-Guillen}
\affiliation{Departamento de F\'isica de Part\'iculas, Universidad de Santiago de Compostela and Instituto Galego de F\'isica de Altas Enerxias (IGFAE) E-15782 Santiago de Compostela, Spain}
\author{A. Wereszczynski}
\affiliation{Institute of Physics,  Jagiellonian University,
Reymonta 4, Krak\'{o}w, Poland}
\begin{abstract} 
The BPS Skyrme model is a specific subclass  of Skyrme-type field theories
which possesses both a BPS bound and infinitely many soliton solutions
(skyrmions) saturating that bound, a property that makes the model a very
convenient first approximation to the study of some properties of nuclei
and hadrons.  A related property, the existence of a large group of
symmetry transformations, allows for solutions of rather general shapes,
among which some of them will be relevant to the description of physical
nuclei.  

\noindent We study here the classical symmetries of the BPS Skyrme model,
applying them to construct soliton solutions with some prescribed shapes,
what constitutes a further important step for the reliable application of
the model to strong interaction physics.  
\end{abstract}

\maketitle 

%%%%%%%%%%%%%%%%%%%%%%%%%%%%%%%%%%%%%%%%%%%%%%%%%%%%%%%%%%%%%%
%%%%%%%%%%%%%%%%%%%%%%%%%%%%%%%%%%%%%%%%%%%%%%%%%%%%%%%%%%%%%%
\section{Introduction}\label{sec:introduction}
The Skyrme model~\cite{skyrme} (SM),  a non-linear field theory for an
$SU(2)$-valued field, is meant to be a low energy effective theory,
describing some interesting aspects of strong interaction physics. In this
model, pions play the role of primary fields (excitations around the
trivial vacuum), whereas nucleons and nuclei are, on the other hand,
represented by topological solitons, collective
excitations which are part of the nonperturbative spectrum of the theory. 

The application of the SM to nuclear and hadronic physics has
been quite successful at a qualitative level~\cite{AdNaWi}-\cite{man1}, but
it encounters some problems once a more detailed, quantitative agreement, is
required. The main obstacle for this is the absence of (almost) BPS
solutions in the original SM, as well as in its standard generalizations.
Indeed, although there exists a BPS bound already in the original model, as
proposed by Skyrme, nontrivial soliton solutions cannot saturate this
bound. As a consequence, higher solitons, meant to describe larger nuclei,
are strongly bound, in striking contrast to the weak binding energies of
physical nuclei. 

Some alternatives approaches to improve this situation have been recently
advanced. Basically they imply the extension of the symmetries of the Skyrme type theory to conformal transformations \cite{Sut1} or to volume preserving diffeomorphisms \cite{BPS-Sk}. It it the aim of this paper to further
elaborate on one of them, namely, the proposal of~\cite{BPS-Sk}.  

The SM may be generalized in a rather straightforward way, by
simply adding some judiciously chosen extra terms to its defining
Lagrangian~\cite{jackson}-\cite{ding}.  Indeed, the addition of extra terms
becomes a quite natural step when one recalls the fact that the SM is an
effective theory, supplemented with the condiment of some simplicity and symmetry
constraints. 
In fact, assuming, as one usually does, that we want to maintain the
field content of the original model,  as well as its Poincar\'e invariance and
the standard Hamiltonian interpretation (Lagrangian quadratic in time
derivatives), the number of possible terms is in fact quite restricted. 
One may then just have: a potential term (no derivatives), a standard kinetic term (the nonlinear
sigma model term) quadratic in first derivatives, the `Skyrme term'
originally introduced by Skyrme (quartic in derivatives) and,
finally, a term which is the square of the baryon number current
(topological current), which is sextic in derivatives. 

As it has been demonstrated in~\cite{BPS-Sk}, there is a submodel, termed
`BPS Skyrme model' (BPSSM), defined by a Lagrangian consisting of
just the potential and sextic terms, which satisfies some quite interesting
properties. Indeed, it possesses a BPS bound, and infinitely many BPS
solutions saturating this bound.  Besides, it has been also shown
in~\cite{BPS-Sk} that the static energy functional of the model is invariant under 
an infinite number of symmetry transformations, a fact that is obviously
related to the properties enunciated in the previous sentence. 

Among the symmetry transformations, an interesting type are the volume
preserving diffeomorphisms (VPDs), since they  are precisely the symmetries
of an incompressible fluid, a fact pointing to a possible relation to the
liquid drop model for nuclei. The BPSSM, therefore, has several appealing
features from the point of view of the description of nuclei (see, for
example, \cite{BPS-Sk}, \cite{BoMa}). The model is, in fact, constructed assuming that the coherent (topological) excitations play an especially important role in strong interaction physics. This assumption is directly related to the suppression of the usual kinetic term in the Lagrangian and, as a consequence, one might expect that the BPSSM will not lead to reliable results in the weak field regime. To overcome these shortcomings, it may be necessary to augment the lagrangian by further structures for a more consistent description
of nuclear or hadron physics. There are, for instance, initial data for
which the BPSSM does not have a well-defined Cauchy problem; thus, a
standard kinetic term must either be added explicitly or induced by quantum
corrections, to remedy this situation. 

We think that the BPSSM provides an approximation which may be quite
reliable for the study of static properties and for the dynamics in a
region of relatively high density (i.e., with a not too small baryon charge
density) like, e.g., in a soliton background.  On the other hand, it will
generally not be reliable in near-vacuum regions, and moreover cannot be
applied at all to consider perturbative phenomena corresponding to quantum
fluctuations of the pion field around the vacuum, since the
dominant term would then be non-quadratic.  

Because of the above, it would be important to relate the properties of
solutions of the BPSSM to the corresponding solutions of more general
Skyrme-type models.  A stumbling block which immediately pops up when
attempting this task is the different sizes of the respective
spaces of solutions, which are in turn due to the different symmetry groups
of the models. The solutions of the BPSSM may have almost any symmetry, due
to the huge symmetry group
of the field equations. In particular, there are spherically symmetric
solutions (i.e., with spherically symmetric energy densities) for all the
possible values of the baryon charge, $Q_B$. 
This is not the case, on the other hand, for the original SM and its non
BPS generalizations. Typically, the $Q_B =1$ skyrmion is spherically
symmetric, the $Q_B=2$ skyrmion has cylindrical symmetry, while higher-charge
skyrmions have, at most, a set of discrete symmetries. Indeed, their energy
densities are invariant under some discrete subgroup of the rotation group
SO(3) (see, for example, \cite{BaSu1}-\cite{man-sut-book}).  
A skyrmion of the BPSSM with the same set of discrete symmetries would,
therefore, be a good starting point for the inclusion of physical effects induced by adding
other extra terms to the Lagrangian.  Because of this, it would be
important to find a method for the systematic construction of solutions of the BPS
Skyrme model with some prescribed symmetries. 

It is the purpose of the present notes to investigate the space of BPS
solutions further, making explicit use of its symmetries as a tool to
generate new solutions. To that end, we shall take the spherically
symmetric ones as a starting point for the construction.  Finally, we shall
show that all local solutions may, in fact, be
constructed in this way.

This article is organized as follows: In section~\ref{sec:themodel}, we
define the model, and introduce our notation and conventions. Then we
construct the classical Hamiltonian in section~\ref{sec:hamilton}. The BPS
bound is considered in~\ref{sec:bps}.  In~\ref{sec:symmetries} we explore
the issue of symmetries, within both the Lagrangian and Hamiltonian
contexts. In~\ref{sec:solutions} we derive and discuss the main properties
of the BPS solutions of the model. We also construct several explicit
classes of solutions with some prescribed symmetries, including the important case of discrete symmetries.
In~\ref{sec:conclusions} we summarize our results and conclusions.

%%%%%%%%%%%%%%%%%%%%%%%%%%%%%%%%%%%%%%%%%%%%%%%%%%%%%%%%%%%%%
%%%%%%%%%%%%%%%%%%%%%%%%%%%%%%%%%%%%%%%%%%%%%%%%%%%%%%%%%%%%%
%%%%%%%%%%%%%%%%%%%%%%%%%%%%%%%%%%%%%%%%%%%%%%%%%%%%%%%%%%%%%
\section{The model}\label{sec:themodel}
The Lagrangian density ${\mathcal L}$, which has an $SU(2)$ valued
field $U$ as dynamical variable, may be written as follows:
\begin{equation}\label{eq:defl06}
	{\mathcal L} \,=\, {\mathcal L}_{06}\,=\,-\lambda^2 \pi^2 \, B_\mu
	B^\mu - \mu^2 {\mathcal V}(U,U^{\dagger}) \;, 
\end{equation}
where $\lambda$ is a positive constant, $B^\mu$ denotes the topological
current:
\begin{equation}
B^\mu \,=\, \frac{1}{24 \pi^2} \;\epsilon^{\mu\nu\rho\sigma} {\rm
tr}\big(L_\nu L_\rho L_\sigma) \;,\;\;\;
L_\mu \equiv U^\dagger \partial_\mu U \;,
\end{equation}
and ${\mathcal V}$ is a potential density. The current $B^\mu$ is
`topologically conserved', namely, it can be shown to be conserved,
regardless of the equations of motion. 
The resulting conserved charge, $Q_B$, is therefore given by:
\begin{eqnarray}\label{eq:defQ}
	Q_B &=& \int d^3x \, B_0 \;=\;\frac{1}{24 \pi^2}  \int d^3x \,
	\epsilon^{ijk} {\rm tr}\big(L_i L_j L_k) \nonumber\\
	 &=& \frac{1}{24 \pi^2} \, \int d^3x \,
	\epsilon^{ijk} \, {\rm tr}\big(
	U^\dagger \partial_i U  
	U^\dagger \partial_j U  
	U^\dagger \partial_k U \big) \;,
\end{eqnarray}
the degree of the map ${\mathbb R}^3 \to S^3$, an integer which is
invariant under arbitrary globally well-defined coordinate transformations, as well as under 
global isospin rotations of $U$.  It is, in fact, invariant under the much bigger group of target space transformations leaving invariant a certain target space volume form, see below.

To proceed to the classical equations of motion, it is convenient to
introduce a specific parametrization for the three degrees of freedom of
$U$.

Following~\cite{BPS-Sk} , we use a real scalar field $\xi$ plus a
3-component unit vector $\mathbf{\hat{n}}$, so that:
\begin{equation}
U(x) \,=\, e^{i \xi(x) \mathbf{\hat{n}}(x) \cdot \mathbf{\tau}} \;,
\end{equation}
where ${\mathbf \tau}$ are the three Pauli matrices. The real scalar $\xi$
runs from $0$ to $\pi$, while the two independent parameters defining
$\mathbf{\hat{n}}$ may be taken as the two components of a complex variable
$u$, by means of a stereographic projection: 
\begin{equation}\label{eq:stereodef}
	\mathbf{\hat{n}}=\frac{1}{1+|u|^2} \left( u+\bar{u}, -i ( u-\bar{u}),
|u|^2-1 \right) \;.
\end{equation}
In this way, one obtains for the Lagrangian density an expression in terms
of $\xi$, $u$ and $\bar u$:
\begin{equation}
	{\mathcal L}\,=\, \frac{  \lambda^2 \sin^4 \xi}{(1+|u|^2)^4} \;\left(  \epsilon^{\mu 
\nu \rho \sigma} \xi_{\nu} u_{\rho} \bar{u}_{\sigma} \right)^2
-\mu^2 {\mathcal V}(\xi)
\end{equation}
where the lower indices in those variables denote partial derivatives with
respect to the spatial coordinates, and we have assumed that the potential 
may only depend on $U$ through \mbox{$\mbox{tr}\,U$}.

With the notation ${\mathcal V}_\xi \equiv \partial_\xi {\mathcal V}$, the
Euler--Lagrange equations read:
\begin{eqnarray}\label{eq:eleq}
\frac{\lambda^2 \sin^2 \xi}{(1+|u|^2)^4} \partial_{\mu} ( \sin^2 \xi \; 
H^{\mu}) + \mu^2 {\mathcal V}_{\xi}=0 \nonumber\\
\partial_{\mu} \left( \frac{K^{\mu}}{(1+|u|^2)^2} \right)=0 \;,
\end{eqnarray}
where 
$$ H_{\mu} = \frac{\partial (  \epsilon^{\alpha \nu \rho \sigma} \xi_{\nu} 
u_{\rho} \bar{u}_{\sigma})^2}{ \partial \xi^{\mu}}, \;\;\; K_{\mu} = 
\frac{\partial (  \epsilon^{\alpha \nu \rho \sigma} \xi_{\nu} u_{\rho} 
\bar{u}_{\sigma})^2}{\partial \bar{u}^{\mu}}.$$
These objects satisfy, by construction, the relations 
\begin{equation}
H_{\mu} u^{\mu}=H_{\mu} \bar{u}^{\mu}=0, \; K_{\mu}\xi^{\mu}=K_{\mu}
u^{\mu}=0, 
\;\; H_{\mu} \xi^{\mu}=K_{\mu} \bar{u}^{\mu} = 2 (  \epsilon^{\alpha \nu \rho 
\sigma} \xi_{\nu} u_{\rho} \bar{u}_{\sigma})^2 \;, 
\end{equation}
which are often useful. 
%%%%%%%%%%%%%%%%%%%%%%%%%%%%%%%%%%%%%%%%%%%%%%%%%%%%%%%%%%%%%%%%%%%%%%%%%%%%%%%
%%%%%%%%%%%%%%%%%%%%%%%%%%%%%%%%%%%%%%%%%%%%%%%%%%%%%%%%%%%%%%%%%%%%%%%%%%%%%%%
%%%%%%%%%%%%%%%%%%%%%%%%%% Hamiltonian formalism %%%%%%%%%%%%%%%%%%%%%%%%%%%%%%
%%%%%%%%%%%%%%%%%%%%%%%%%%%%%%%%%%%%%%%%%%%%%%%%%%%%%%%%%%%%%%%%%%%%%%%%%%%%%%%
%%%%%%%%%%%%%%%%%%%%%%%%%%%%%%%%%%%%%%%%%%%%%%%%%%%%%%%%%%%%%%%%%%%%%%%%%%%%%%%
\section{Hamiltonian and static energy}\label{sec:hamilton}
In order to construct the Hamiltonian, we first introduce a more compact
notation, in terms of three real fields $\xi^{(a)}$, with $a=1,2,3$, such that
$u = \xi^{(1)} + i \xi^{(2)}$ and $\xi^{(3)} \equiv \xi$.

Then ${\mathcal L}$ may be written as follows ($\xi_0^{(a)} \equiv \partial_0 \xi^{(a)}$):
\begin{equation}
	{\mathcal L} \,=\,\frac{1}{2} \xi_0^{(a)} G_{(ab)} \xi_0^{(b)}
	-\frac{ 4 \lambda^2 \sin^4 \xi^{(3)} 
	\big(\epsilon^{ijk} \xi^{(1)}_i \xi^{(2)}_j
	\xi^{(3)}_k\big)^2  }{\big[1+(\xi^{(1)})^2
	+ (\xi^{(2)})^2 \big]^4} -\mu^2 {\mathcal V}(\xi) \;.
\end{equation}
where the kinetic term is determined by a metric $G_{(ab)}$, given by:
\begin{equation}
	G_{(ab)}\,=\, \frac{ 2 \lambda^2 \sin^4 (\xi^{(3)})}{[ 1 +
	(\xi^{(1)})^2 + (\xi^{(2)})^2]^4} \,
	{\mathcal Q}_i^{(a)} {\mathcal Q}_i^{(b)} 
\end{equation}
where:
\begin{equation}
	{\mathcal Q}_i^{(a)} \,=\, \epsilon_{ijk} \, \epsilon^{a b c}
	\xi_j^{(b)} \xi_k^{(c)} \;.
\end{equation}
In order to see whether the system defined by ${\mathcal L}$ is regular or
not, we note that ${\mathbb Q} \equiv [{\mathcal Q}_i^{(a)}]$ the $3 \times
3$ matrix defined by the nine elements ${\mathcal Q}_i^{(a)}$ ($i=1,2,3$;
$a=1,2,3$) is proportional to the cofactor matrix of the matrix ${\mathbb
X}\equiv [\xi_i^{(a)}]$:
\begin{equation}
	{\mathbb Q} \,=\, 2 \; {\rm cof}(\mathbb X) \;.
\end{equation}
Thus, we see that the metric $[G_{(ab)}]$ (hence, the Lagrangian system) is regular
if and only if $\det[\xi_i^{(a)}] \neq 0$. In other words, the regularity
of the system is equivalent to the non vanishing of the Jacobian determinant:
\begin{equation}\label{eq:defj}
	{\mathcal J} \,\equiv\, \det[{\mathbb X}] 
	\,=\, \det \big[\frac{\partial\xi^{(a)}}{\partial x_i}\big] \neq 0\;,
\end{equation}
for the mapping between the sphere (i.e., one-point compactified $\mathbb{R}^3$) in coordinate space and the one in
$SU(2)$.

Under the assumption that (\ref{eq:defj}) holds true, the inverse of
${\mathbb G}= [G_{(ab)}]$ may be found by elementary algebra. Indeed,
\begin{equation}
	[{\mathbb G}^{-1}]^{(ab)} \,=\, 
	\frac{[1 +(\xi^{(1)})^2 + (\xi^{(2)})^2]^4}{8 \lambda^2 \,
		{\mathcal J}^2 \, \sin^4(\xi^{(3)})} \xi_i^{(a)} \xi_i^{(b)} \;.
\end{equation}

Thus, the Hamiltonian density in terms of the variables $\xi^{(a)}$, its
spatial derivatives, and their canonical momenta $\Pi^{(a)}$, becomes:
\begin{eqnarray}
	{\mathcal H} &=&
	\frac{[1 +(\xi^{(1)})^2 + (\xi^{(2)})^2]^4}{16 \lambda^2 \,
		{\mathcal J}^2 \, \sin^4(\xi^{(3)})}  \Pi^{(a)}\xi_i^{(a)}
		\, \xi_i^{(b)}\Pi^{(b)} \nonumber\\
		&+&\frac{ 4 \lambda^2 \sin^4 \xi^{(3)} 
	\big(\epsilon^{ijk} \xi^{(1)}_i \xi^{(2)}_j
	\xi^{(3)}_k\big)^2  }{\big[1+(\xi^{(1)})^2
	+ (\xi^{(2)})^2 \big]^4} +\mu^2 {\mathcal V}(\xi) \;,
\end{eqnarray}
which, for a Lagrangian like the one we are considering, coincides with the
energy density of the system. In particular, for the static configuration
case to be considered in the forthcoming sections, the total energy
$E$ is:
\begin{equation}\label{eq:statice}
	E \,=\,	\int d^3 x \Big\{ 4 \lambda^2 \frac{\sin^4 \xi^{(3)} 
	\big(\epsilon^{ijk} \xi^{(1)}_i \xi^{(2)}_j
	\xi^{(3)}_k\big)^2  }{\big[1+(\xi^{(1)})^2
+ (\xi^{(2)})^2 \big]^4} +\mu^2 {\mathcal V}(\xi) \Big\} \;.
\end{equation}

We have shown that the regularity of the system depends on the field
configurations considered. Specifically, the system is singular in
regions where the fields take their vacuum values ($\xi^{(a)} =$ const.
such that ${\mathcal V}(\xi^{(3)})=0$). This already demonstrates that,
while the system may provide a good approximation to the description of
static properties of nucleons and nuclei via solitons (Skyrmions) and for
the dynamics in regions with nonzero baryon charge density (where it is
regular by construction), its fully consistent application to dynamical
nuclear physics requires additional structures like, e.g., quantum
corrections, or the inclusion of further terms in the Lagrangian.

%%%%%%%%%%%%%%%%%%%%%%%%%%%%%%%%%%%%%%%%%%%%%%%%%%%%%%%%%%%%%%%%%%%%%%%%%%%%%%%
%%%%%%%%%%%%%%%%%%%%%%%%%%%%%%%%%%%%%%%%%%%%%%%%%%%%%%%%%%%%%%%%%%%%%%%%%%%%%%%
%%%%%%%%%%%%%%%%%%%%%%%%%%%%%%%%% BPS %%%%%%%%%%%%%%%%%%%%%%%%%%%%%%%%%%%%%%%%%
%%%%%%%%%%%%%%%%%%%%%%%%%%%%%%%%%%%%%%%%%%%%%%%%%%%%%%%%%%%%%%%%%%%%%%%%%%%%%%%
%%%%%%%%%%%%%%%%%%%%%%%%%%%%%%%%%%%%%%%%%%%%%%%%%%%%%%%%%%%%%%%%%%%%%%%%%%%%%%%
\section{BPS bound}\label{sec:bps} 
The static energy functional in (\ref{eq:statice}), or, in terms of the variables $\xi$ and $u$ introduced previously,
\begin{equation}
E=\int d^3 x \left[  \frac{\lambda^2 \sin^4 \xi}{(1+|u|^2)^4} 
(\epsilon^{mnl} i \xi_m u_n\bar{u}_l)^2 +\mu^2 {\mathcal V}(\xi) \right]
\end{equation}
obeys a Bogomolny bound. Indeed, 
$$ 
E= \int d^3 x \left( \frac{\lambda \sin^2 \xi}{(1+|u|^2)^2} 
\epsilon^{mnl} i \xi_mu_n\bar{u}_l \pm \mu \sqrt{\mathcal V} \right)^2 \mp \int d^3 x 
\frac{2\mu \lambda \sin^2 \xi \sqrt{{\mathcal V}}}{(1+|u|^2)^2} \epsilon^{mnl} i \xi_m 
u_n \bar{u}_l 
$$
$$ 
\geq  \mp \int d^3 x \frac{2\mu \lambda \sin^2 \xi \sqrt{\mathcal V}}{(1+|u|^2)^2} 
\epsilon^{mnl} i \xi_m u_n \bar{u}_l =
$$
\begin{equation} \label{bobo}
\pm (2\lambda \mu \pi^2 )\left[ \frac{-i}{\pi^2}
\int d^3 x \frac{ \sin^2 \xi \sqrt{\mathcal V}}{(1+|u|^2)^2} 
\epsilon^{mnl}  \xi_m u_n \bar{u}_l \right] 
\equiv 2\lambda \mu \pi^2 \langle \sqrt{\mathcal V} \rangle |B|
\end{equation}
where $\langle \sqrt{\mathcal V}\rangle$ is the average value of $\sqrt{\mathcal V}$ on the target space S$^3$. 
The corresponding Bogomolny (first order) equation is
\begin{equation}
\frac{\lambda \sin^2 \xi}{(1+|u|^2)^2} \epsilon^{mnl} i \xi_mu_n\bar{u}_l 
= \mp \mu \sqrt{\mathcal V} .
\end{equation}
The static second order field equations may be derived from the squared
Bogomolny equation by applying a gradient $\partial_k$ and by projecting
onto $\epsilon_{ijk}\partial_j \xi^{(a)}$ where $\xi^{(a)} \equiv (\xi
,u,\bar u)$. We remark that a completely analogous BPS bound can be found for the BPS baby Skyrme model in one lower dimension \cite{izq}-\cite{Speight1}.

Another interesting observation is that the BPS equation can be formulated in the language of a non-linear generalization of the static (vacuum) Nambu-Poisson equation.  Indeed the left hand side can be recast into the Nambu-Poisson three-bracket \cite{nambu}
\begin{equation}
\left\{ X^A, X^B, X^C \right\} = \epsilon^{mnl} \frac{\partial X^A}{\partial x^m}\frac{\partial X^B}{\partial x^n}\frac{\partial X^C}{\partial x^l}
\end{equation}
where the target space embedding coordinates $X^A, A=1,2,3,4$ form a three-sphere $S^3$ (i.e., $(X^A)^2=1$) and are related to the previous coordinates like $X^a = n^a \sin \xi $, $a=1,2,3$, and $X^4 = \cos \xi$. Then, the generalized Nambu-Poisson dynamics is given by 
\begin{equation} \label{gen-NPeq}
\frac{d X^A}{d t} =\epsilon^{ABCD} \left\{ X^B, X^C, X^D \right\}  + X^A \sqrt{\mathcal{V}(X^4)} ,
\end{equation}
which differs from the standard case by the additional factor $\sqrt{\mathcal{V}}$ in the last term \cite{nambu}. Obviously, although the dynamics of the BPS Skyrme model is profoundly different, the BPS equation provides static solutions to this generalized Nambu-Poisson equation. Such solutions may be interpreted as vacuum configurations of the underlying hyper-membrane Lagrangian \cite{fairlie}. We remark that if one assumes from the outset that the target space variables $X^A$ span a three-sphere, as we do in this paper, then there is no dynamics in Eq. (\ref{gen-NPeq}), i.e., $\frac{ dX^A}{dt}=0$, as follows from the fact that the r.h.s. of (\ref{gen-NPeq}) is proportional to $X^A$ in this case. This just corresponds to the well-known result that the static vacuum equations for the hyper-membrane imply that the brane embedding coordinates $X^A$ span a three-sphere \cite{fairlie}. So, our model generalizes the static hyper-membrane action, 
with a correspondence between the BPS solitons and
the vacuum membrane configurations, but with completely different dynamics.

It may be instructive to compare the BPS bound arising above with a $1+1$
dimensional analogue: the search for (non-trivial) static minimum energy configurations 
for the Sine-Gordon model.  Here, a real scalar field $\varphi$ is in the
presence  of a potential density ${\mathcal U}(\varphi)$ which allows for
non-trivial topology. The Lagrangian density is:
\begin{equation}
	{\mathcal L} \,=\, 
	\frac{1}{2} (\partial_\mu \varphi)^2 - {\mathcal U}(\varphi)
\end{equation}
\begin{equation}
{\mathcal U}(\varphi)\,=\,\frac{m^4}{\lambda} \big[ 1 -
\cos(\frac{\sqrt{\lambda}}{m}\varphi)\big] \;.
\end{equation}
The static energy is then:
\begin{equation}
	E_\varphi \,=\,\int_{-\infty}^{+\infty} dx_1 \big[\frac{1}{2}
	(\partial_1\varphi)^2  + {\mathcal U}(\varphi)\big] \;.
\end{equation}
The non-negative potential  has non-trivial minima for
\begin{equation}
	\varphi = \varphi_N = \frac{2 \pi m}{\sqrt{\lambda}} N \;,\;\; N
	\in {\mathbb Z} \;,  
\end{equation}
all of them having zero energy. Finite energy vacuum configurations must
tend to one of the minima when $x_1 \to \pm \infty$.

The topologically conserved current is
$j^\mu = \frac{\sqrt{\lambda}}{2\pi m}\epsilon^{\mu\nu} \partial_\nu \varphi$ 
($\mu, \nu = 1, 2$), which obviously satisfies $\partial \cdot j =0$. 
Its associated topological charge is quantized:
\begin{equation}
	Q_\varphi \,=\,  \frac{\sqrt{\lambda}}{2\pi m}\, 
	\int_{-\infty}^{+\infty} dx_1 \partial_1 \varphi(x) \;=\; N \;,
\end{equation}
it is a constant of motion, and it is akin to a winding number, if one
interprets $\varphi$ as an angular variable. 

Note the striking similarity with the BPS Skyrme model, when one writes the
energy as follows:
\begin{equation}
	E_\varphi \,=\,\int_{-\infty}^{+\infty} dx_1 \big[\frac{1}{2}
	( \frac{2 \pi m}{\lambda})^2  (j_0)^2  + {\mathcal U}(\varphi)\big] \;.
\end{equation}

The static energy then also verifies a Bogomolny-like bound, since:
\begin{equation}
	E_\varphi \,=\,\int_{-\infty}^{+\infty} dx_1 
	\big[\frac{1}{\sqrt{2}} \frac{d\varphi(x_1)}{dx_1}  \pm
	\sqrt{{\mathcal U}(\varphi)}\big]^2  
	\mp \sqrt{2} \int_{-\infty}^{+\infty} dx_1 \frac{d\varphi(x_1)}{dx_1} 
	\sqrt{{\mathcal U}(\varphi)}  \;.
\end{equation}
Thus:
\begin{eqnarray}
	E_\varphi & \geq & \pm \,\sqrt{2}\,\int_{-\infty}^{+\infty} dx_1 \frac{d\varphi(x_1)}{dx_1} 
	\sqrt{{\mathcal U}(\varphi)} \nonumber\\
	&=& \pm \sqrt{2} 
	\,
	\frac{2\pi m}{\sqrt{\lambda}} \;
	|\langle {\mathcal U} \rangle |	
	\,
	|Q_\varphi| \nonumber\\
	&=&  2 \, \sqrt{2} \, \pi \,\frac{m^3}{\lambda} \, |Q_\varphi| \;. 
\end{eqnarray}
where:
\begin{equation}
	|\langle {\mathcal U} \rangle |	 \,=\, \frac{1}{\varphi_1 -
	\varphi_0} \,  
	\int_{\varphi_0}^{\varphi_1} d\varphi
	\sqrt{{\mathcal U}(\varphi)} \;,  
\end{equation}
the average of $\sqrt{{\mathcal U}(\varphi)}$ over the fundamental region.

Of course, the first order equations that result from saturating the bound
may be found by other methods; they lead to the well-known static solutions
by a single quadrature. What we learn from the comparison with this model
is that the particular form of the Lagrangian of the BPS Skyrme model
involving the square of the topological current, is what makes it produce
quite powerful constraints on the solution. It is interesting to note that the kinetic term in this 1+1 dimensional example allows for two different interpretations, either as a standard kinetic term or as the topological current squared, which is no longer true in higher dimensions. In other words, the simple Sine-Gordon type soliton model in 1+1 dimensions allows for two different generalizations to higher dimensions, generalizing either the standard kinetic term or the topological current, and the model studied in the present paper just corresponds to the second case.

%%%%%%%%%%%%%%%%%%%%%%%%%%%%%%%%%
%%%%%%%%%%%%%%%%%%%%% SYMMETRIES %%%%%%%%
%%%%%%%%%%%%%%%%%%%%%%%%%%%%%%%%%%%%%%
\section{Symmetries}\label{sec:symmetries}
The Lagrangian certainly has the standard Poincar\'e symmetries. Besides, the
sextic term is the square of the pull back of the target space volume form
on S$^3$,
\begin{equation}
dV= -i\frac{\sin^2 \xi}{(1+|u|^2)^2}d\xi du d\bar u
\end{equation}
so this sextic term is invariant under target space diffeos which do not
change this form (the volume preserving diffeos (VPDs) on S$^3$). The
potential only depends on $\xi$, so it is still invariant under those
diffeomorphisms which do not change $\xi$, i.e., under the diffeos which obey
\begin{equation}
\xi \to \xi \, , \quad u\to \tilde u(u,\bar u, \xi) \, ,\quad 
(1+|\tilde u|^2)^{-2} d\xi d\tilde u d\bar{\tilde u} = 
(1+| u|^2)^{-2} d\xi  u d\bar{ u} .
\nonumber
\end{equation}
The symmetries mentioned so far are symmetries of the action, i.e. Noether symmetries.  

The static energy functional has some further symmetries. Indeed, it is
invariant under volume preserving diffeos on the base space $\mathbb{R}^3$,
as can be seen easily. The Bogomolny equation has even more symmetries as
we want to demonstrate now. For this purpose we introduce the new target
space coordinates
$$
u=ge^{i\Phi}=\tan (\chi/2) e^{i\Phi}, \quad H(g)=\frac{1}{1+g^2}, 
$$
(for later convenience we also introduced $\chi$, which together with $\xi$ and $\Phi$ provides the standard hyperspherical coordinates on the target S$^3$), and
\begin{equation} \label{F-eq}
 F(\xi)= \frac{\lambda}{\mu}\int d\xi \frac{\sin^2 \xi}{\sqrt{{\mathcal V}(\xi)}}
\end{equation}
and rewrite the Bogomolny equation as
\begin{equation}
\nabla F(\xi) \cdot \nabla H(g) \times \nabla \Phi =\pm 1
\end{equation}
or, in terms of differential forms
\begin{equation} \label{mod-Bog-2}
dFdHd\phi = \pm dx^1dx^2dx^3
\end{equation}
from which it is obvious that the Bogomolny equation has as its symmetries
all the VPDs both in base space and in a modified target space defined by
the volume form $dFdHd\Phi$. The above equation implies, in fact, that all
local VPDs on base space produce local solutions of the BPS equation. The
problem is that, in general, a local solution cannot be extended to a
global one, because of the different geometry and topology of the base
space and the modified target space. The modified target space is defined
by the volume form
\begin{equation}
dFdHd\Phi = -\frac{\lambda}{\mu} \frac{\sin^2 \xi}{\sqrt{{\mathcal V}(\xi)}}\sin \chi d\xi d\chi d\Phi
\end{equation}
and differs from the volume form on S$^3$ by the additional factor
$1/\sqrt{\mathcal V}$. There does not exist a unique riemannian metric
giving rise to this volume form, but a natural choice which assumes that
the S$^2$ spanned by $u$ (i.e., $\chi$ and $\Phi$) remains intact is
\begin{equation}
ds^2 = d\xi^2 + \frac{\sin^2 \xi }{\sqrt{{\mathcal V}(\xi)}}(d\chi^2 + \sin^2 \chi d\Phi^2 ).
\end{equation} 
For ${\mathcal V}=1$ this is just the round metric on S$^3$ in
hyperspherical coordinates, but for nontrivial potentials the resulting
target space manifold is different. Indeed, potentials which may support
finite energy skyrmion solutions must have vacua $\xi = \xi_0$ where
${\mathcal V}(\xi_0 )=0$, and 
the above metric is singular at the vacuum values $\xi_0$. These
singularities may either be integrable (i.e., the function $F$ defined in
(\ref{F-eq}) is well-defined and finite even at vacuum values $\xi
=\xi_0$), in which case the total volume of the modified target space is
still finite. In the opposite case, the total volume is infinite. One
further conclusion may be drawn immediately by integrating Eq.
(\ref{mod-Bog-2}). If the total volume of the modified target space is
finite, then any skyrmion solution of the BPS equation must have compact support (i.e., be a "compacton").
Further, its volume must be equal to $|B|$ times the total volume of the
modified target space, where $B$ is the winding number. For equivalent
results for the case of the BPS baby Skyrme model in one lower dimension,
we refer to \cite{Speight1}. 
\\
We remark that for $\mathcal{V} = \sin^4 \xi$ the metric on the target space describes in fact a 3 dimensional cylinder with a very simple skyrmion solution (see below).
%%%%%%%%%%%
%%%%%%%%%%%%
\section{Solutions}\label{sec:solutions}
%%%%%%%%%%%%%
As already said, locally, any VPD on base space will provide a solution of
the BPS equation, but this solution will, in general, not be extendible to a
global, genuine one (i.e., a skyrmion), because of the nontrivial topology
one should have on the modified target space. 
A more promising strategy is the following: start from a simple known
solution which may follow from a simple ansatz. Then one may generate new
solutions by composing the given solution with a VPD on base space $\mathbb{R}^3$. 
If the VPD is well-defined on the whole of $\mathbb{R}^3$, then it will map
genuine skyrmions into genuine skyrmions. In the case of compactons, we may
even relax this condition, since it is then sufficient for the VPD on base
space to be well-defined in the region of the compacton.  

To proceed, let us first find some simple solutions with the help of an
ansatz in spherical polar coordinates
\begin{equation}
\xi = \xi (r), \quad \chi = \chi (\theta) , \quad \Phi = n\varphi
\end{equation}
which inserted into the BPS equation yields:
\begin{equation}
-\frac{\lambda}{\mu} \frac{\sin^2 \xi}{\sqrt{{\mathcal V}(\xi)}}\sin \chi
d\xi d\chi d\Phi = \mp r^2 \sin \theta drd\theta d\varphi \;,
\end{equation}
leading to $\chi =\theta$ and 
$$
-\frac{n\lambda}{\mu} \frac{\sin^2 \xi}{\sqrt{\mathcal V}} d\xi = \mp r^2 dr
$$
or, after the coordinate transformation:
\begin{equation}
y= \frac{\mu}{3\sqrt{2}\lambda n}r^3
\end{equation}
to the autonomous ODE:
\begin{equation}
\frac{\sin^2\xi}{\sqrt{2{\mathcal V}(\xi)}} \xi_y = - 1 \;.
\end{equation}
We have chosen the sign which leads to a negative $\xi_y$, which is
compatible with the boundary conditions $\xi (r=0)=\pi$, $\xi (r=\infty)=0$
for a potential which takes its vacuum at $\xi_0 =0$. 

Let us consider now the symmetries of these solutions. This issue depends 
on the criterion used to characterize that symmetry. Note that a given solution
will not be invariant under any rotation, because it depends on the two angular
coordinates $\theta$ and $\varphi$. The energy density, on the other hand,
depends only on the radial coordinate $r$ and is, therefore, spherically symmetric.
 Note, however, that there exists another symmetry criterion, often used
for solitons, whereby there is spherical symmetry when the effect of a base
space rotation on a solution can be undone by a corresponding target space rotation. 
Under this criterion, only the solution with topological charge $n=1$ is spherically 
symmetric (i.e., all rotations can be undone). Solutions with higher winding number $n$
only have cylindrical symmetry, i.e., only a rotation about the $z$ axis
$\varphi \to \varphi + \alpha$ can be undone by a target space rotation (a
phase transformation $u \to e^{-in\alpha} u$). 

In any case, we shall call all solutions of the spherically symmetric
ansatz "spherically symmetric solutions" in what follows.  We shall first review some general properties of these spherically symmetric solutions and, in a next step, construct solutions with lesser symmetries. 
%%%%%%%%%%%%%%%%%%%%%%%%%%%%
\subsection{Solutions with spherical symmetry}
%%%%%%%%%%%%%%%%%%%%%%%%%%%%
Many qualitative aspects of solutions maybe easily derived from the particular form of the potential, which should be contrasted with the typical situation in general Skyrme models, where similar results usually require a full three-dimensional numerical simulation.  
\\
First of all, depending on the form of the potential in the vicinity of the vacuum, one can distinguish three types of solitonic configurations: compactons (where the solution approaches its vacuum value at a strictly finite distance) and exponentially as well as power-like localized solutions. Using the BPS equation and expanding the potential at a vacuum (e.g., at $\xi=0$),  $\mathcal{V}=\mathcal{V}_0\xi^\alpha+...$, one  easily finds that for $\alpha <6$ one gets compactons. There is also one exponentially localized solution for $\alpha=6$, while for $\alpha >6$ we find power-like localized solitons. 
\\
Another important feature of solutions reflects the number of vacua of the potential. It is easy to prove  that for one-vacuum potentials the BPS solutions are of the nucleus type (no empty regions in the interior), while two-vacuum potentials lead to shell-like configurations.    
\\
Let us present some particular examples. For the most elaborated family of one vacuum potentials, the so-called old potentials
\begin{equation}
 \mathcal{V}_{old}=\left(\mbox{Tr} \left(\frac{1-U}{2} \right) \right)^a \;\; \rightarrow 
\;\; \mathcal{V}(\xi)=(1- \cos \xi)^a 
\end{equation}
(where $a$ is a real positive parameter), we find (besides the previously known compacton) a solution with exponential tail ($a=3$) in implicit form
$$ \cos \frac{\xi}{2}+\ln \tan \frac{\xi}{4} = - \frac{y}{2}$$,
and power-like localized solutions.  E.g., for $a=6$ we get
$$ \xi =2\; \mbox{arc cot} \sqrt[3]{3\sqrt{2}y}. $$ 
A family of two-vacuum potentials is given by  
\begin{equation}
 \mathcal{V}_{shell\; I}=\left(\mbox{Tr} \left(\frac{1-U}{2} \right) \mbox{Tr} \left(
\frac{1+U}{2} \right) \right)^a\;\; \rightarrow \;\; V(\xi)=(1- 
\cos^2 \xi)^a,
\end{equation} 
which is the chiral counterpart of the so-called new baby potential.  The vacua exactly coincide with the boundary values for the scalar field i.e., $\xi=0, \pi$. From the BPS property of the solution one can immediately see that the energy density should have a shell structure with two zeros: one at the center of the soliton, while the second (outer zero) can be located at a finite distance (compact shells) or approached asymptotically at infinity. Without losing generality (the potential is symmetric under the change of the vacua) we assume that $\xi=0$ is the outer vacuum. Of course, the inner vacuum can only be reached at a finite point as $y \geq 0$. This implies that only compact solitonic shells are acceptable. Specific examples of exact solutions are, for $a=1$
$$
\xi = \left\{
\begin{array}{lc}
 \arccos (\sqrt{2} y -1) & y \in \left[0,\sqrt{2} \right] \\
0 & y \geq \sqrt{2}, 
\end{array} \right. 
$$
and for $a=2$
$$
\xi = \left\{
\begin{array}{lc}
\pi -\sqrt{2}y & y \in \left[0,\frac{\pi}{\sqrt{2}} \right] \\
0 & y \geq \frac{\pi}{\sqrt{2}}.
\end{array} \right.
$$
The latter solution is, in fact, a solution for the case when the target space is a three-dimensional cylinder, as $\sin^2 \xi / \sqrt{\mathcal{V}} =const.$ 
\\
In order to deal with non-compact shell skyrmions, we need to modify our potential in such a way that one vacuum (say, the inner vacuum at $\xi=\pi$) is always approached in a compacton manner . A simple choice is
\begin{equation}
 \mathcal{V}_{shell\;  II}=\mbox{Tr} \left(\frac{1+U}{2} \right) \left( \mbox{Tr} \left(
\frac{1-U}{2} \right) \right)^a\;\; \rightarrow \;\; V(\xi)=(1+ \cos \xi ) (1-
\cos \xi)^a
\end{equation} 
Again, we find compact shell skyrmions $a<3$
 $$ 
 \xi = \left\{
\begin{array}{lc}
\arccos \left[  1 - \left( 2^{\frac{3-a}{2}}-\frac{3-a}{\sqrt{2}} y \right)^{\frac{2}{3-a}} \right] & y \leq \frac{\sqrt{2}}{3-a}  \\
0 & z \geq \frac{\sqrt{2}}{3-a} 
\end{array} \right.
$$
an exponentially localized skyrmion for $a=3$
$$\xi =
\xi = \arccos \left[ 1-2e^{-\sqrt{2}y}\right] ,$$
and shell skyrmions which extend to infinity but are localized in a power-like manner ($a>3$)
$$\xi =
\arccos \left[  1 - \left( 2^{\frac{3-a}{2}}+\frac{a-3}{\sqrt{2}} y \right)^{\frac{2}{3-a}} \right] .$$

%%%%%%%%
\subsection{Solutions with cylindrical symmetry}
%%%%%%%%
Now we assume that a spherically symmetric solution has been found, and we
want to use symmetry transformations to map them to new solutions. In a
first step we construct solutions with cylindrical symmetry, using the
ansatz (in cylindrical coordinates)
\begin{equation}
\xi = \xi (\rho ,z) ,\quad g = g(\rho ,z) , \quad \Phi =n \varphi
\end{equation}
where $\rho^2 = (x^1)^2 + (x^2)^2 $, $z=x^3$.  The Bogomolny equation for this ansatz may be written like
\begin{equation}
dF^{(n)}dH = \pm dqdp
\end{equation}
where $F^{(n)} =nF$ and
$$
q=\frac{\rho^2}{2} ,\quad p=z
$$
or like the Poisson bracket
\begin{equation}
\{F^{(n)},H\} \equiv \frac{\partial F^{(n)}}{\partial q} \frac{\partial H}{\partial q} - \frac{\partial F^{(n)}}{\partial p} \frac{\partial H}{\partial q} =\pm 1.
\end{equation}
Further, we know that it has the spherically symmetric solution 
\begin{equation}
g=g_s = \tan (\theta/2) = \frac{\rho}{\sqrt{\rho^2 + z^2} +z} = \frac{\sqrt{2q}}{\sqrt{2q + p^2} +p} \equiv g_s(q,p)
\end{equation}
and (depending on the potential)
\begin{equation}
\xi = \xi_s (r) = \xi_s (\sqrt{2q + p^2}) \equiv \xi_s(q,p).
\end{equation}
As a consequence, a general solution with spherical symmetry may be written like
\begin{equation}
\xi (q, p) = \xi_s (Q(q,p),P(q,p)) , \quad g(q,p)= g_s (Q(q,p),P(q,p))
\end{equation}
where $(Q,P)$ are related to $(q,p)$ via a canonical transformation, i.e., $\{Q,P\} =1$. 

A first class of examples is given by $$Q=U(q) ,\quad P=\frac{p}{U'(q)} $$
where $U'(q)\neq 0 \; \forall \; q$ must hold. Further, it should hold that $\lim_{q\to 0} U(q)/q = const.$ to have a well-behaved function near $\rho =0$. Among these examples the scale transformation $Q=a^2 q, P= a^{-2}p$ can be found, which corresponds to the scale transformation $x^1 \to ax^1$, $x^2 \to ax^2$ and $x^3 \to a^{-2} x^3$.
Another class of examples is
$$Q=\frac{q}{U'(p)} ,\quad P=U(p). $$ 
%%%%%%%%
\subsection{Solutions with discrete symmetries}
%%%%%%%%
Here, we want to construct a class of base space VPDs which transform
solutions with spherical or cylindrical symmetry into solutions which only
preserve symmetries w.r.t. to some discrete rotations about the $z$ axis.
Concretely, we want to consider solutions which may be written like
\begin{equation}
\xi = \xi (\rho ,z) = \xi_s (\tilde \rho ,z) ,\quad g = g(\rho ,z) =g_s (\tilde \rho ,z) , \quad \Phi =n \tilde\varphi
\end{equation}
where $\xi_s$, $g_s$, $\Phi =n\varphi$ constitute a known solution with
either spherical or cylindrical symmetry. That is to say, we consider base
space VPDs which act nontrivially only on $\rho$ and $\varphi$, where for
simplicity we restrict ourselves to the following transformations,
\begin{equation}
\tilde \rho =\tilde \rho (\rho ,\varphi ), \quad \tilde \varphi = \tilde \varphi (\varphi) .
\end{equation}
Using $q=\rho^2 /2$ as before, and $\tilde q = \tilde q (q,\varphi)$, the condition for the transformation to be a VPD simplifies to
\begin{equation}
d\tilde q d\tilde \varphi = dqd\varphi .
\end{equation}   
A class of formal solutions is given by
\begin{eqnarray}
\tilde q &=& (f')^{-1} q \nonumber \\
\tilde \varphi &=& f(\varphi ) 
\end{eqnarray}
in close analogy to the results of the last section. In order to define
genuine diffeomorphisms, however, the transformations have to obey some
further conditions.  
In particular, for the new coordinates $\tilde q$ and $\tilde \varphi$ to
define polar coordinates on $\mathbb{R}^2$ they must satisfy the boundary
conditions
\begin{eqnarray}
\tilde q (q=0,\varphi )=0, &&\tilde q(q=\infty, \varphi )=\infty , \nonumber \\
 \tilde \varphi (\varphi =0)=0, && 
\tilde \varphi (\varphi =2\pi) =2\pi .
\end{eqnarray}
In addition, the vector field generating the flow induced by the coordinate
transformation must be well-defined (nonzero and nonsingular) on the whole
of $\mathbb{R}^2$. A class of examples fulfilling all the required conditions
is given by $f=\varphi + (c/m)\sin m\varphi $, i.e., by the class of
transformations
\begin{eqnarray}
\tilde q &=& \left( 1+c \cos m \varphi \right)^{-1} q \qquad \qquad m \in \mathbb{N} \nonumber \\
\tilde \varphi &=& \varphi + \frac{c}{m} \sin m \varphi \qquad \qquad c \in \mathbb{R}, \quad |c|<1.
\end{eqnarray} 
Clearly, if a solution $\xi^{(a)}_s (\rho ,z,\varphi )$ is invariant under
rotations about the $z$ axis (in the sense that its energy density
is invariant under these rotations), then the new solution $\xi^{(a)}_s
(\tilde \rho ,z,\tilde \varphi )$ is invariant only under the discrete set
of rotations $\varphi \to \varphi + (2\pi /m)$. 
%%%%%%%%%%%%%%%%%%
%%%%%%%%%%% SUMMARY %%%%%%%
%%%%%%%%%%%%%%%%%%%%
\section{Summary}\label{sec:conclusions}
We explored in detail the symmetries of the static energy functional of the
BPSSM, and of its related BPS equation. Then we applied these symmetries
to the systematic construction of new solutions, starting from known ones.
This is in the spirit of the dressing methods of classical
integrability \cite{dress}, which is an open problem for  higher
dimensional generalizations \cite{nos}, an initial motiviation of
this work.
Specifically, this allowed us to construct solutions with some prescribed
symmetries, what is quite relevant to the physical problem one wants to
consider. 
We gave concrete examples of solutions with cylindrical symmetry and with
symmetries w.r.t. some discrete subgroup of the group $SO(2)$ of rotations
about the $z$ axis. In this context, it would be interesting to construct
solutions with the symmetries of platonic bodies or other discrete
subgroups of the full rotation group $SO(3)$ (crystallographic groups),
because solitons with these symmetries frequently show up as true
minimizers of the energy in the original Skyrme model or some of its
generalizations \cite{BaSu1}-\cite{man-sut-book}. The corresponding
volume-preserving diffeomorphisms producing solutions with these symmetries
will be more complicated than the ones constructed in the present paper,
and it almost certainly  will be more difficult to find them. 
\\
This issue is under current investigation. 

\section*{Acknowledgements}
C.A., J.M.Q., J.S.-G. and A.W. acknowledge financial support from the Ministry of Education, Culture and Sports, Spain (grant FPA2008-01177), the Xunta de Galicia (grant INCITE09.296.035PR and Conselleria de Educacion), the Spanish Consolider-Ingenio 2010 Programme CPAN (CSD2007-00042), and FEDER.
C.D.F has been supported by CONICET, CNEA, and UNCuyo. Further, AW was supported by polish NCN grant 2011/01/B/ST2/00464.
J.S.-G. thanks V.A. Rubakov for interesting suggestions.

\end{document}